\documentclass[12pt,fullpage]{article}

\newcommand{\be}{\begin{equation}}
\newcommand{\ee}{\end{equation}}
\newcommand{\beq}{\begin{eqnarray}}
\newcommand{\eeq}{\end{eqnarray}}
\newcommand{\AmS}{{\protect\the\textfont2
\thispagestyle{empty}

A\kern-.1667em\lower.5ex\hbox{M}\kern-.125emS}}

\advance\voffset by -1cm
\advance\hoffset by -.5cm
     
\begin{document}
\begin{flushright}
{TIFR/TH/06-19} \\
{July  2006} \\
\end{flushright}

\vspace{1em}

\begin{center}
{\large{Albert Einstein: His Annus Mirabilis 1905$^*$}} \\ [6mm]
{\sf Virendra Singh} \\ 
INSA C.V. Raman Research Professor \\ [3mm]
Department of Theoretical Physics \\
Tata Institute of Fundamental Research \\
1, Homi Bhabha Road, Mumbai 400 005, India
\end{center}

\vspace{2em}

\abstract{
Einstein in 1905, his year of miracles, wrote five papers which mark a
watershed between classical physics and modern physics. They 
dealt with problem of reality of atoms, theory of special relativity
which overthrew Newtonian conceptions of space and time, and his 
revolutionary light quantum hypothesis which together with
Planck's work on black body radiation started the quantum
revolution. We put these discoveries in the context of that period and
also indicate their later influence.}

\vfill

\hrule width 5cm 
\medskip

{\noindent$^*$Inaugural Lecture of ``Einstein Lecture Series'' (9 lectures)
delivered at Nehru Center, Mumbai on July 9, 2005. Similar lectures
were also given at other institutions.}

\newpage

\section{Year of Physics (2005)}

We are having a worldwide celebration of physics throughout this year
(2005). The year 2005 has been declared the ``Year of Physics'' by
UNESCO. It is the centenary year of the ``Annus Mirabilis'', ie the
Miracle year, 1905, of Albert Einstein. During this year he published
a set of five papers dealing with the existence of atoms, special
relativity including the now famous equation $E = MC^2$ expressing the
equivalence of energy content $E$ and inertial mass $M$ of a body, as
well as on the quantum theory together with its application to the
photoelectron effect. These papers mark the watershed between
classical physics of Isaac Newton, Michael Faraday and John Clerk
Maxwell and the modern physics. It is therefore entirely appropriate
that this centenary year of the annus mirabilis be celebrated as the year
of physics.

We may also mention that the decade of 1895 to 1905 was extremely rich
in discoveries which established the existence of a number of
phenomenon which were not explicable within classical physics. This
crisis would require for its resolution a change of classical
framework to that of modern physics involving relativity and quantum
theory. In 1895, Roentgen discovered X-rays. In 1896 the radioactivity
was discovered by Henri Becquerel and magnetic field effect on
spectral lines by Pietr Zeeman. In 1897, J.J. Thompson established the
existence of electrons. In 1900, Max Planck introduced quantum ideas
in physics. As has been noted in this connection, the twentieth
century in physics began not in the year 1900, but a full five years
earlier in 1895. It is also fitting that this fruitful decade
(1895--1905) was capped by the Annus Mirabilis of Einstein.

Before we proceed to discuss in detail why the year 1905 is referred
to as the annus mirabilis, let us briefly recall an earlier year, 1666
which is also referred to by the same designation. It was the annus
mirabilis of Isaac Newton, who came to symbolise the emergence of not
only classical physics and astronomy but classical mathematics. 

\section{Annus Mirabilis of Isaac Newton} 

\subsection{John Dryden}

The noted restoration English poet, John Dryden, published a long
poem, some 116 pages in print, in 1688 usually referred to briefly as
``Annus Mirabilis''. It is worth quoting the title in full which was
``Annus Mirabilis, The Year of Wonders, MDCLXVI, An Historical
Poem. Also A Poem on the Happy Restoration and Return of His Late
Sacred Majesty, Charles the Second. Likewise A Panegyric on His
Coronation Together With a Poem to My Lord Chancellor, Presented on
New-Year Day 1662. And a Elegy on the Death of Kind Charles the Second
By John Dryden, Esq''. It celebrated the annus mirabilis, 1666, the
year of wonders and some of the wonders so celebrated were survival of
London from the Great Fire and the Victory of English fleet over the 
Dutch. 

This the how the term ``Annus Mirabilis'' entered the English
language. The term was soon appropriated by scientists to refer to the
achievements of Isaac Newton during that year, 1666, which truly could
be called ``wonders'' unlike ``wonders'' celebrated by John Dryden in
his poem. 

\subsection{Isaac Newton}

Since some of Newton's discoveries took place in the previous year, ie
1665, we should more properly speak of his Anni Mirabilis, ie years of
wonders, 1665-1666. Newton was born at the village of Woolsthorpe on
christmas day 1642.  Coincidentally it was the year in which Galileo
died. Newton was admitted to Trinity College at the Cambridge
University at the age of eighteen.  Soon after he had completed his
Bachelor's degree, the University was closed down due to a serious
epidemic of the ``Great Plagues'' at Cambridge in June 1665. All the
professors and students went home and Newton returned to his ancestral
village of Woolsthorpe to live with his grandmother and mother. During
this period of enforced idleness Newton started thinking about
mathematics and scientific problem and had one of those rare
concentrated burst of creativity which resulted in his laying the
foundation of classical mathematics, physics and astronomy.

Some fifty year later Newton himself wrote an account of period for
des Maizeaux. This account is almost invariably quoted in this
connection and is as follows: ``In the beginning of the year 1665 I
found the Method of approximating series and the Rule for Reducing any
dignity of any Binomial into such a series. The same year in May I
found the method of Tangents of Gregory and Slusius, and in November
had the direct method of fluxions and the next year in January had the
Theory of Colours and in May following I had entrance into ye inverse
method of fluxions. And the same year I began to think of gravity
extending to ye orb of the Moon and $\cdots$ I deduced that the forces
wch keep the Planets in their Orbs must be reciprocally as the squares
of their distances from the centers about which they revolve $\cdots$
All this was in the two plague years of 1665-1666. For in those days I
was in the prime of my age for invention and minded Mathematics and
Philosophy more than at any other time since''.

Note that Newton was a fresh undergraduate and only twentyfour years
of age at the time when he discovered analysis, both differential and
integral calculus, theory of colours and the law of universal
gravitation.  

\section{Annus Mirabilis: Albert Einstein}

\subsection{Early life}

Albert Einstein was born on 14 March 1879 at Ulm in Germany. In his
childhood, in 1889, his father had presented him a pocket magnetic
compass and young Einstein found the behavior of magnetic needle,
which always point in a fixed north south direction, as a profound,
almost mystical, experience. This gave him the idea that the physical
world is subject to laws. 

On July 28, 1900 he was granted Diploma by ETH (Eidgen\"ossische
Technische Hochschule) at Zurich. His grades in different courses
were: Theoretical Physics - 5/6, Experimental Physics - 5/6, Astronomy
- 5/6, Theory of Functions - 5.5/6 and the Diploma Paper -- 4.5/6. An
overall grade of 5 our of a total of 6 is quite good and might come as
a surprise to those who have been exposed to the idea that Einstein
was a poor student.  He was indeed a poor student in his elementary
school, which was more due to bad teaching, but had clearly remedied
the situation by the time he was at ETH, Zurich. Not only that towards
the end of that year, on December 13, 1700, he sent his first research
paper for publication to the well known journal ``Annalen der Physik''. 

He started work as a clerk at the Patents office, Bern on probation at
a salary of 3500 swiss francs per year. It seems that his father had
to use his personal influence with one of his friends for Albert to
get this job. Soon afterwords his father died on October 10, 1902 and
thus one of sources of financial security for him was cut off. On top
of that his expenses increased when he married his fellow student
Mileva Maric and with the birth of their first son Hans Albert on
March 14, 1904, Luckily he was confirmed in his job at the Patent
office on September 16, 1904. 

During this period (1901-1904) when he had no proper academic
appointment and subject as well to financial insecurities, it is
surprising that he published five papers in Annalen der Physik. We
need not comment on first two of these which deal with capillary
phenomenon. But the remaining three, on his discovery of the ensemble
method is statistical mechanics were rather important. Here, however,
he had been scooped by Josiah W. Gibbs, who had obtained these results
some what earlier. 

\subsection{Einstein in 1905}

During this year, Einstein published five papers on statistical
physics, special theory of relativity and the quantum theory apart
from completing his Ph.D. dissertation. 

In chronological order these were as follows:
\begin{enumerate}
\item[{(1)}] Light quantum paper: The paper ``On a heuristic point of
view concerning the production and transformation of light'' was
received by Annalen der Physik on March 18, 1905. This was published
in Annalen der Physik 17, 132-148 (1905). 
\item[{(2)}] Thesis on Molecular Sizes: The Ph.D. dissertation ``On a
new determination of the Molecular Dimensions'' was completed on April
30, 1905. It was printed at Bern and submitted to University of
Z\"urich on July 20, 1905. He also sent a paper based on the thesis to
Annalen der Physik soon after the thesis was accepted on August 19,
1905 by the University which appeared in Annalen der Physik 19,
289-305 (1906) next year. 
\item[{(3)}] Brownian Motion paper: The paper ``On the motion of small
particles suspended in liquids at rest required the Molecular Kinetic
theory of heat'' was received on May 11, 1905 for publication and
appeared in Annalen der Physik, 17, 549-560 (1905).
\item[{(4)}] Special theory of relativity paper: The paper ``On the
electrodynamics of moving bodies'' was received for publication on
June 30, 1905 and appeared in Annalen der Physik 17, 891-921 (1905). 
\item[{(5)}] $E=mc^2$ paper: The paper ``Does the inertia of a body
depend on its energy context'' was received for publication on
September 27, 1905 and appeared as Annalen der Physics 18, 639-641
(1905). 
\end{enumerate}

Einstein, besides the above, sent another paper on Brownian motion on
December 19, 1905 to Annalen der Physik which was published next
year. 

We had mentioned earlier Newton's own account of his Anni Mirabilis
some half a century after the event. We have an account by Einstein of
his work in his annus mirabilis which was written during that very
year. It occurs in letters he wrote to his friend Conrad Habicht. He,
together with Einstein and Maurice Solovine, was a member of the
triumvirate `Olympia Academy', who used to meet regularly in evenings
to have wide ranging intellectual discussions extending from
philosophy to physics. 

Einstein wrote to Habicht on May 18 or 25, 1905: ``$\cdots$ But why
have you still not sent me your dissertation? $\cdots$ I promise you
four papers in return, the first of which I might send you soon, since
I will soon get complimentary reprints. The paper deals with radiation
and the energy properties of light and is very revolutionary as you
will see if you send me your work \underbar{first}. The second paper
is a determination of the true sizes of atoms from the diffusion and
the viscosity of dilute solutions of neutral substances. The third
proves that, on the assumption of molecular theory of heat, bodies on
the order of megnitude 1/1000 m.m., suspended in the liquids, must
already perform an observable random motion that is produced by the
thermal motion; in fact physiologists have observed (unexplained)
motions of suspended small, inanimate, bodies, whose motion they
designate as ``Brownian Molecular motion''. The fourth paper is only a
rough draft at this point, and is an electrodynamics of moving bodies
which employes a modification of the theory of space and time; the
purely kinematical part of this paper will surely interest
you. $\cdots$''. 

Einstein again wrote to him on some Friday during the period June 30,
1905 -- September 22, 1905 to bring him uptodate with his later work
as follows: ``$\cdots$ A consequence of the study on electrodynamics
did cross my mind. Namely, the relativity principle, in association
with Maxwell's fundamental equations, requires that the mass be a
direct measure of the energy contained in a body; light carriers mass
with it. A noticeable reduction of mass would have to take place in
the case of radium. The consideration is amusing and seductive; but
for all I know, God Almighty might be laughing at the whole matter and
might have been leading me around by the nose''. 

We have excised the purely personal remarks and banter from these
letters. 

In the rest of the writeup we shall now discuss these contribution in
somewhat more detail and provide their background and context so as to
appreciate them more properly. We shall not follow the chronological
order in which they were written but rather the order in which they
make a transition from classical physics to modern
physics. Chronologically his light quantum paper is first during 1905
but as Einstein himself remarked it is the most revolutionary. The
order followed would therefore be as follows: 
\begin{enumerate}
\item[{1.}] Thesis on molecular motion
\item[{2.}] Brownian motion paper
\item[{3.}] Special theory of relativity and $E=mc^2$ papers
\item[{4.}] Light Quantum paper.
\end{enumerate}

\section{Thesis on Molecular Sizes}

\subsection{The first attempt}

The Ph.D. thesis which Einstein wrote in 1905 was not his first
attempt at submitting a thesis for this degree. He first submitted a
Ph.D. dissertation in November 1901. It is not known as to what the
topic was. It is also not clear as to why Einstein withdrew it soon in
February 1902. As he wrote to his friend Michele Besso from Bern on a
Thursday (January 22(?) 1903): ``I have recently decided to join the
ranks of Privatdozenten, assuming, of course that I can carry through
with it. On the other hand I will not go for a doctorate, because it
would of little help to me, and the whole comedy has bcome
boring''. He however changed his mind about a doctorate degree soon
afterwards. 

\subsection{Thesis on molecular sizes}

Till 1909 ETH was not recognized as an institution allowed to grant
doctoral degrees. However under a special arrangement the ETH students
were permitted to submit their doctoral dissertations to University of
Z\"urich. Einstein's thesis advisor was Alfred Kleiner who was an
experimental physicist, specialised in instrumentation. He however
also had a broader interest in basic physics. The thesis was dedicated to
his fried Marcel Grossman, who would eventually help him with tensor
calculus in his formulation of general theory of relativity in 1915.

Einstein motivated the thesis topic by pointing out that though there
have been many determinations of molecular sizes till that date they
all have used kinetic theory of gases. His proposed method would be
the first to use phenomenon in liquids. Even though not exactly the
first to do so, it was indeed the first one to give results comparable
in accuracy from those obtained by much more developed kinetic theory
of gases. This is qite remarkable in the absence of any available
kinetic thoery of liquids. Einstein had chosen his thesis problem
totally on his own as was acknowledged by his thesis advisor Kleiner. 

\subsection{Main results in the thesis}

The thesis has two main results: 
\begin{enumerate}
\item[{(i)}] \underbar{\sf Viscosity of dilute solutions} \\
Let $\eta$ and $\eta_s$ be respectively the viscosity of the solvent
and solution made of a solute in this liquid. Einstein's
hydrodynamical investigation led to the result
\[
\eta_s = \eta \left[1 + \left({5\over 2}\right)^* \varphi \right]
\]
where the $\varphi$ is the fraction of volume occupied by the solute
molecules in the solution. The solute volume fraction $\varphi$ is
given by
\[
\varphi = {4\pi \over 3} (N_A a^3) (\rho_s/m_s)
\]
where $N_A$ is the Avodgadro number, $a$ the radius of the solute
molecules, $\rho_s$ (and $m_s$) refer to the massdensity (and the
molecular weight) of the solute. We have put a star on the numerical
factor $5/2$ above as this factor was missed out in Einstein
dissertation. 

Einstein assumed that the solution is a dilute one and further that
the solute molecules do not dissociate in the solution.  He proceeds to
calculate the change in flow of the solvent around solute particles,
taken to be spheres, and shows that it results in an effective
visiosity coefficient $\eta_s$ for the solution as given above. 

\item[{(ii)}] \underbar{\sf Diffusion coefficient of solute
molecules}\\
Einstein also established a formula for diffusion coefficient $D$ of
the solute particle in the solvent liquid. He showed that
\[
D = RT/(6\pi\eta N_A a) 
\]
where $R$ is the universal gas constant and $T$ is the absolute
temperature of the liquid. 

We thus see that a measurement of the relative change in viscosity of
a solution due to a solute allows us to determine the product $N_A
a^3$ while a measurement of the diffusion coefficient allows us to
know the product $N_A a$.  Together these two measurements allow us a
determination of both the Avogadro Number, $N_A$, and the size of the
solute molecules, $a$. Einstein applied his analysis to available data
on dilute solutions of sugar molecules in water and obtained
\beq
a &=& 9.9 \times 10^{-8} cm , \nonumber \\
N_A &=& 2.1 \times 10^{23}. \nonumber
\eeq
On using improved data, which became available soon he obtained a
better value of $N_A$ given by
\[
N_A = 4.15 \times 10^{23}
\]
during 1906. This was first of the three methods proposed by Einstein
during 1905 for determining $N_A$. This clearly points to the
importance Einstein attached to the problem of determining the
Avogadro number. 
\end{enumerate}

\subsection{Comments on the dissertation}

As Alfred Kleiner noted in his report, dated 22-23 July, 1905, on the
thesis: ``The arguments and calculations to be carried out are among
the most difficult ones in hydrodynamics, and only a person processing
perspicacity and training in the handling of mathematical and physical
problems could dere to tackle them, and it seems to me that
Mr. Eination has proved that he is capable of working successfully on
scientific problems; I would therefore recommend that the dissertation
be accepted''. He however added ``since the main achievement of
Einstein's thesis consists in the handling of differential equatons,
and hence is mathematical in character and belongs in the domain of
analytical mechanics, I would like to ask the dean also to approach my
colleague Professor Burkhardt (Heinrich Burkhardt was Professor of 
Mathematics at University of Zurich) for an expert opinion''. 

His expert opinion was as follows: ``At the request of my colleague,
Professor Kleiner, I reviewed the dissertation of Mr. Einstein and
checked the most important part of his calculations, that is, all of
the places indicated by Professor Kleiner. What I checked, I found to
be correct without exception, and the manner of treatment demonstrate
{\it a thorough command of the mathematical methods, involved
$\cdots$}''. Incidently the correct factor of $5/2$, in the expression
for $\eta_s$ was again missed out in this checking. In view of the
discrepency between the experimental results on the solution viscosity
of Jacques Bancelin, working in the laboratory of Jean Perrin,
Einstein requested his student and collaborator, Ludwig Hopf, to check
his calculations again. Hopf was successful in finally finding the
missing factor of $5/2$ in Einstein's expression. Hopf's correction
was communicated to Perrin by Einstein on Jan 12, 1911. If this
correction is used then we get the much more satisfactory value
\[
N_A = 6.56 \times 10^{23} .
\]

Initially this dissertation was foreshadowed by other papers of
Einstein during this year. However, eventually, this is the paper of
Einstein which has received the highest citation in view of it's use
by molecular physicists and chemists. May be citation index is not
such an infallible guide to the significance of a paper!

\section{Brownian Motion}

\subsection{Atomic theory around the end of nineteenth century}

Modern chemistry dates back to John Dalton's book ``New System of
Chemical Philosophy'' in 1808 in which he proposed his system of a
finite number of chemical elements. All the molecules were taken as
composed of atoms of these chemical elements. Amedeo Avogadro in 1811
proposed that, under conditions of equal temperature and pressure,
equal volumes of gases contain the same number of molecules for all
gases. This number for a mole of gas was named by Jean Perrin as the
Avogadro Number $N_A$. Avogadro's law presupposes the reality of
molecules. Most chemists, however, used atomic theory in the
nineteenth century as a theoretical heuristic device to bring order
into the description of chemical phenomenon. They did not necessarily
subscribe to their reality. 

In the second half of the nineteenth century, the Kinetic theory of
gases, which posited the gases to consist of moving molecules, made
rapid progress. Clausius, in 1857, suggested that heat is a form of
moleculer motion. John Clerk Maxwell proposed his famous distribution
law for the molecular velocities in a gas. Ludwig Boltzmann gave his
equation which set out to reduce all thermodynamic phenomenon to
mechanical description using molecules. These developments in the
kinetic theory of gases gave a big boost to the atoms being real
entities. 

At the end of nineteenth century most physicists and chemists thus
either believed in the reality of molecules or at least were willing
to use them as heuristics. In view of the fact that all the evidence
for the atoms was indirect, as atoms were not directly seen, there was
still a small but powerful opposition to the idea of their
reality. The great physical chemist Ostwald, as well as George Helm,
regarded atoms to be mathematical constructs. The situation in regard
to atoms was similar to that of ``quarks'' as constituents of matter
in twentieth century. Ostwald had his own program, `Energetics', in
which the prime ontological entity was energy. Max Planck also was of
that persuasion at that time since he regarded laws of thermodynamics
to be absolute laws. While in Boltzmann's atomic view the second law
of thermodynamics, regarding entropy, was only statistical in nature
and not absolute. Even the great physicist and philosopher Ernst Mach
was opposed to the reality of the atoms in view of ``positivist''
slant of his philosophy. In 1905 Einstein made a decisive impact on
this debate through his paper on Brownian Motion. 

\subsection{Einstein's contribution}

In his paper on Brownian motion, Einstein investigated the random
motions executed by visible, but very small, particles in a
liquid. The visible random motion of these particles was taken to
arise from their being buffeted by the incessent motion of the
invisible liquid molecules. His hope was that such a study would be
convincing enough about the reality of the underlying molecules of the
liquid. As he noted ``It will be shown in this paper that according to
molecular -- kinetic theory of heat, bodies of a microscopically
visible size suspended in liquids must, as a result of thermal
molecular motion, perform motions of such magnitude that they can
easily be detected by a microscope''. He continues ``It is possible
that the motions to be discussed here are identical with the so-called
``Brownian molecular motion''; however, the data available to me on
the latter are so imprecise that I could not form a definite opinion
on this matter''. 

Robert Brown, the english botanist, had observed random motion of
pollen grains in a liquid in 1828. The motion was analogous to a
drunkard's walk around a lamp post. Brown as a result of his
experiments ruled out the possibility that the observed motion was due
to pollen grains being moved by some vital force ie due to their
living nature. Many different suggestions such as effect of
capillarity, role of convection currents, evaporation, interaction
with light, and electrical forces were put forward to explain these
random motion. Even kinetic theory was proposed as a possible
explanation but Von Nageli, in 1879, ruled it out for reasons which
appeared cogent. He took straight segments on the path of a Brownian
particle to be their free motion between two collisions with
molecules. We now know that even these straight segments arises due to
the effect of multiple collisions with atoms. In fact one of
achievements of Einstein in this paper was to clarify the physically
significant observations to make on these particles. 

Einstein calculated the diffusion constant $D$ for the suspended
microscopic particles, of the size `a' of the order of one-thousandath
of a millimeter, in the liquid and showed that it is given by
\[
D = RT/(6 \pi \eta a N_A)
\]
where $T$ is the temperature, $\eta$ the viscosity of the liquid. As
before $N_A$ is the Avogadro Number and a determination of $D$ would
provide us another method to measure it. One would recall that same
formal expression for $D$ had appeared in his thesis on molecular
sizes. There it was however for solute molecules while here it is for
suspended particles in the liquids. One would think that these two
situations being analogous, except for the size of the diffusing
particle, the same formula should be valid. However in those days it
was not believed that Vant-Hoft's law of osmotic pressure is
applicable for both the solute molecules as well as for suspended
particles. Einstein showed using molecular kinetic theory that it is
indeed valid for both.  

Einstein then showed that a measurement of mean square fluctuation in
the $x$-component of position of a Brownian particle, $<x^2>$ in time
$t$ provides us with a way to measure $D$ as
\[
\langle x^2 \rangle = 2 Dt ,
\]
assuming that this prediction of $<x^2>/t$ being constant is correct. That
can however be always checked by the experiment. This is the first
example of a fluctuation -- dissipation theorem in physics. 

\subsection{Jean Perrin}

J. Perrin, and his group, beginning 1908, carried out a series of
beautiful experiment on the Brownian of colloid particles in
suspension. They were able to produce the colloidal particles of a
uniform size. Their work resulted in a complete confirmation of
Einstein's results and a precise determination of the Avogadro's
number. As a result of Perrin's work the atomism triumphed. Even the
arch-disbeliever in atoms, F.W. Ostwald, was convinced of their
existence. As he wrote in a new edition of his ``Outline of
Chemistry'', published by the end of first world war, ``I am now
convinced that we have recently become possessed of experimental
evidence of the discrete or grained nature of matter for which the
atomic hypothesis sought in vain for hundreds and thousands of
years''. As Perrin, himself, wrote in his book Les Atomes (1913),
``The atomic theory has triumphed. Until recently still numerous, its
adversaries, at last overcome, now renounce their misgivings, which
were for so long, both legitimate and undeniably useful''. J. Perin
was awarded the Nobel Prize for this work in 1926. 

\subsection{Further work}

Einstein also gave another derivation of the diffusion equation in
this paper based on treating the motion of a Brownian particle as, to
use modern terminology, a random Markov process. This derivation is to
be contrasted to the classical derivations which were based on continuum
mechanics. He thus connected diffusion process of many particles,
approximated as a continuum, to the random-walk problem of individual
particles. Further developments in the theory of Brownian motion have
resulted in great progress in the study of stochastic processes,
fluctuation phenomenon thus giving birth to most of statistical
physics. All these have their origins in the Brownian motion paper of
Einstein. 

\section{Classical Physics and It's Discontents}

The thesis and the Brownian motion paper of Einstein were of great
importance to physics in view of their bearings on the ``reality of
molecules'', as well as for other reasons mentioned earlier. They were
however perfectly in the mold of classical physics. With his papers on
special theory of relativity and on quanta, which we discuss later, he
was launching revolutions in classical physics. To discuss these
conceptual revolutions it is first necessary to give some idea of the
conceptual structure of classical physics. 

\subsection{Classical physics}

The foundations of classical Newtonian dynamics were laid by Isaac
Newton in his annus mirabilis and published in his magnum opus
``Principia'', or to give it its' full title ``Philosophiae Naturalis
Principia Mathematica'', in 1687. Newtonian world consists of point
particles which influence each other by mutual forces ``acting at a
distance from each other''. He also discovered the universal inverse
square law of gravitation between masspoints, thus unifying physics
and astronomy. The masspoints move with time in an arena of three
dimensional space. Space and time are \underbar{absolute} in the sense
that the motion of the particles does not affect them. Thus the drama
of particle motion is played on an unchanging fixed stage of the space
and time. 

Newton also viewed light as a stream of discrete particles. Christian
Huygen, was first to propose in 1678 that light is better described as
a wave motion. Later discoveries of interference of light by Thomas
Young in 1801, and of diffraction of light by Augustin Fresnel gave a
strong support to the wave theory of light and it was firmly
established. Since it was inconcievable in those days to think of
wavemotion without a medium, which would oscillate and support it's
propagation, a universal medium ``luminiferous aether'' was postuated
to exist. 

The concept of continuous field, which pervades over space, like a
magnetic field, unlike point particles of Newton was introduced by
Michael Faraday around the middle of the nineteenth century. Clerk
Maxwell achieved, in 1864, the synthesis of two disparate fields,
electric and magnetic fields, into a coherent unified field
``electromagnetic field'' in which the two affected each other. A
completely unexpected prediction of Maxwell's equations was that of
transverse electromagnetic waves. The velocity of these waves, now
denoted by $c$, involved electrical and magnetic quantities. On
calculation this velocity $c$ was found to be the same as the known
velocity of light. Maxwell therefore made the brilliant suggestion
that light is the same entity as these electromagnetic waves. This
ends over lightening review of classical physics as it was at the end
of the nineteenth century. 

\subsection{Two clouds on the horizon}

Lord Kelvin, in a very perceptive and insightful lecture before the
Royal Institution in April 1900 talked about two ``Nineteenth century
clouds over the dynamical theory of heat and light''. One of these
involved the continued unsuccessful attempts to experimentally measure
the motion of the earth through luminiferrous aether. The other one of
these referred to the failure of equipartition of energy in classical
statistical mechanics. 

Rest of the Einstein's work during the miracule year 1905 is devoted
to a dispelling of these two ominous clouds on the horizon over the
classical physics. His papers on special theory of relativity deal
with a resolution of ``earth's velocity through aether'' puzzle. This
involves a complete overhaul of classical concepts of space and
time. His paper on the ``light quantum'' deals with other cloud and
ushered in the quantum revolution. We now turn to these papers now. 

\section{Special Theory of Relativity}

\subsection{Galilean relativity}

Newton's laws of motion are valid in a set of special frames of
reference. These are called ``inertial frames of reference''. For
example Newton's first law says that a mass point, not acted upon by any
external force, keeps moving in a straight line with a uniform
speed. Now a particle which is moving in such a fashion in an
earth-laboratory will not appear to moving in a straight line when
viewed from the Sun due to earth's daily rotation and it's annual
revolution around the Sun. Clearly the two frames of reference,
i.e. one in which the earth is at rest and other one in which the Sun
is at rest, can not both be inertial frames of reference. 

How do we know if some particular frame of reference is inertial? We
first note that if a frame of reference $S$ is inertial then any other
frame of reference $S'$ which is moving in a straight line with
uniform velocity, is also inertial. This specifies the class of frames
of reference which are inertial and in which Newton's three laws of
motion hold. In order to characterise the class of the inertial frame
we have to specify at least one of them. In practice for solar system
applications, it was taken to be the frame in which the center of mass
of the solar system is at rest or in uniform rectilinear
motion. Within the accuracy required in these calculation, it was same
as the one in which the center of mass of the universe was at rest of
uniform linear motion or the one in which the system of fixed stars
was at rest or uniform linear motion. 

The rules for comparing the space and time coordinate measurement in
different inertial frames are known as Galilean
transformation. Newton's laws obey the principles of Galilean
relativity. Their form is invariant, ie unchanged, under Galilean
transformation between two inertial reference systems. 

\subsection{Maxwell's electromagnetic theory and Galilean relativity}

Note that as long as Newton's laws of motion are the only fundamental
laws of physics there is no
way in which one can determine the absolute velocity of any inertial
frame with respect to some absolutely fixed point at rest.  This
situation radically changes with the advent of Maxwell's equations for
electromagnetism.

We note that Maxwell's equation do not have the same form in different
inertial frames connected by Galilean transformations.  That is they
are not invariant under them.  For example the velocity of
electromagnetic waves (i.e. light) is a constant.  One can ask, in
which inertial frame is it so?  Because it can not be so in all
inertial frames, which are connected by Galilean relation.  If it is
given by $\vec c$ in its' direction of propagation $S$, it would be
$\vec c + \vec v$ in the frame $S'$ which is moving with a velocity
$\vec v$ rectilinearly with respect to $S$.  The velocity of light was
thus a fixed constant $c$ only in the frame in which the luminiferous
aether is at rest.

Taking advantage of this clash, between invariance of Newton's laws
and non-invariance of Maxwell's electromagnetic theory of light, thus
opens a way by which the motion of earth, for example, can be
experimentally measured with respect to universal aether.  A large
number of methods were thought for this purpose.  All of them gave
a null result.  Experiments were unable to detect the motion of the
earth through aether.  Most celebrated and accurate experiment devised
for this purpose was by Michelson and Morely in 1887, which also
reported a null result.  As Maxwell summarised in an article in
Encyclopedia Britannica, ``The whole question of the state of the
luminiferous medium near the earth, and of it's connection with gross
matter, is very far as yet from being settled by experiment''.

\subsection{Einstein's Resolution: Special theory of Relativity}

Einstein's resolution of ``earth-aether velocity'' problem was
obtained by a thorough revision of Newtonian concepts of absolute
space and absolute time.  In this revision he was guided by his
analysis of the concept of simultaneity.  If the two events take place
in a single frame of reference, e.g. a railway platform or a uniformly
moving railway train on linear tracks, there is no difficulty in
saying whether the two events are simultaneous in the same single
frame of reference.  If you start thinking about the problem one finds
that the two events which look simultaneous in one frame, say railway
platform, are not so in another relatively moving frame, e.g. that of
a moving train.  This is because the light signals used to observe the
two events, whose simultaneity we are discussing, will take different
times in the frames of two relatively moving observers.  This is due
to speed of light signal being finite.  Since simultaneity is not an
invariant concept it follows that time can not be absolute.

Einstein wished to hold on to what is now known as the two postulates
of his ``special relativity theory''.  These are 
\begin{enumerate}
\item[{(i)}] All physical laws have the same form in all inertial
frames i.e. frames of references which move rectilinearly with a
constant velocity with respect to each other;
\item[{(ii)}] The velocity of light is same in all inertial frames.
\end{enumerate}

\noindent These two postulates look irreconcilable within Newtonian
notions of absolute space and absolute time and Galilean
transformation.  However if Newtonian space and time concepts are
modified so as to be in accord with what one has learnt from
Einstein's analysis of `simultaneity', then it was Einstein's insight
that the two postulates considered above can indeed be reconciled.
They can then support a new structure of space and time, now called
space-time.  As Minkowski said in 1908 in a lecture given at Cologne
in 1908 ``Hereafter space and time are bound to fade away and only a
union of the two will preserve an independent reality''.

It is clear since time is not invariant in different inertial frames
we can not maintain the Galilean transformations as the proper ones to
connect them.  Einstein goes on to show that they have to replaced by
Lorentz transformations.  This results in providing purely kinematic
derivation of Fitzgerald-Lorentz length contraction which had been
postulated earlier to explain null aether-earth velocity results.  It
also leads to Einstein time dilation i.e. moving objects live longer.

Under Lorentz transformations Maxwell's theory, it is satisfying to
note, is invariant.  Newtonian dynamics, being invariant under Galilean
transformation, is however not so and it has therefore to be
modified.  These result in mass variation with velocity.  All these
phenomenon are observed in high energy accelerators on a routine
basis. 

Einstein was asked to donate his original manuscript of ``special
relativity paper'' for Kansas War-bond rally for auction.  Since he
did not have it any more he copied out the published version in long-hand.  He
did not however make any corrections despite feeling that he could
have phrased it better in many place.  It went for six and a half
million dollars at the auction and was deposited in the Library of
Congress at Washington.

\subsection{$E = mc^2$}

In his special relativity paper Einstein missed out on a profound
result which led to energy-mass equivalence.  This was published
separately in a 3 page.  Befor this paper one had two seperate
conservation laws, one for the mass and another for energy, in
physical transformations.  With this insight they merged into only one
conservation law.

He noted that ``The mass of a body is a measure of its energy
contrast; If the energy changes by $E$, the mass changes in the same
sense by $E/c^2$, $\cdots$''.  We have modified the notation in this
translation to accord with modern usage.  The published paper has $L$
in place of $E$ and ``$L/9.10^{20}$, if the energy is measured in ergs
and mass in grams'' in place of modern $E/c^2$.

He also noted ``Perhaps it will prove possible to test this theory
using bodies whose energy content is variable to high degree
(e.g. salts of radium)''.  This was quite prophetic in view of its'
eventual sad use in nuclear explosions.  Coming back to his letter to
Habicht, the Lord was indeed chuckling and leading him by the nose.

Incidently the equation $E = mc^2$ is the only equation which occurs
in Bartlett's familier quotation.  It has acquired in our culture an
iconic status and is seen on bill boards, T-shirts and so on.

\subsection{Einstein's later related work}

We just give a chronological listing of important land marks.

\noindent 1907: Discovery of Principle of Equivalence

\noindent 1912-13: Gave metric tensor description of gravitation

\noindent 1915, Nov. 25: Completes his formulation of General theory
of Relativity.  Space-time which was so far regarded as flat and
Euclidean is now modified to a curved Riemannian space time.  The
curvature of space-time is identified with gravitation.

\noindent 1917: First paper on Cosmology

\noindent 1919, May 29: The total solar eclipse expedition led by
British astronomer Eddington confirms General theory of Relativity of
German Einstein.  This provides a shining example of international
peaceful scientific collaboration between scientists, even though
belonging to enemy nations in just concluded First World War.
Einstein becomes a World-icon.

\noindent After 1922: Einstein works unsuccessfully on unifying
electromagnetsm with gravitation.

We also wish to note down about the first ever English translation of
Einstein and Minkowski's papers from German was published by Calcutta
University in 1920.  The translators were M.N. Saha and S.N. Bose and a
historical introduction was provided by P.C. Mahalarobis.

\section{Quantum Revolution}

We now discuss his light quantum paper which was his most
revolutionary one in 1905.

\subsection{The Black Body Radiation: Kirchhoff to Planck}

The origins of the quantum revolution are in the problem of Black-Body
Radiation.  All heated bodies emit radiation energy as well as absorb
it.  A consideration of thermodynamic equilibrium led Gustov Kirchhoff
of Berlin, in 1859 to conclude that the ratio of emissivity to 
absorptivity of the radiation does not depend on the nature of the
heated body.  This ratio, a universal function is the same as the
emissivity of a perfect black body i.e. a body which completely
absorbs all the radiation falling on it.  It was also shown that the
radiation inside a heated cavity is same as black body radiation.  Max
Planck occupied Kirchhoff's chair at Berlin in 1889.  He argued that
as the ratio is independent of the nature of the cavity material he
should be able to calculate it by using a simple model for the
material of the cavity.  The model he used was that it is made of
Hertzian oscillators each with a single frequency $\nu$.  Using this
model he could show that the universal function is related to average
energy of each oscillator of frequency $\nu$ at the temperature $T$ of
the Black Body radiation.  He had this result on 18 May 1899.

If Planck had known the equipartition theorem of classical statistical
mechanics, for average energy, at this point, he would have obtained
the law of Black body radiation now known as Rayleigh-Jeans radiation
law as it was given by Rayleigh in June 1900 and corrected for a
missing factor of 8 by Jeans in June 1905.  Indeed this was first done
by Einstein in his light quantum paper.  Amusingly he did it before
Jeans.  Rayleigh-Jeans radiation law was found applicable only at
small values of $\nu/T$ and not for large values of $\nu/T$.  One thus
became aware of the second cloud on the horizon of classical physics
referred to by Lord Kelvin viz the failure of equipartition of energy.

Guided by the precision experimental results on black body radiation
Planck announced an empirical radiation law on Oct.19, 1900 which
fitted the data perfectly.  The Planck's radiation law is now known to
be the correct law of black body radiation.  It had the same form as
Rayleigh's law for small $\nu/T$ and the form of empirically proposed
Wien's law, given in 1894, for large $\nu/T$.  Planck however had no
theoretical basis for his radiation law.

Planck next presented a derivation of his radiation law.  He was so
desperate that he even used Boltzmann's probability interpretation for
entropy.  The derivation was announced to German Physical Society on
Dec. 14, 1900.  The really new element in his derivation was his
assumption that a Hertzian oscillator, of frequency $\nu$, can emit or
absorb radiation only in integral multiples of a basic quantum of
energy $\epsilon$, where $\epsilon = h\nu$.  The constant $h$ is now
known as Planck's constant.  In classical physics there was no such
discreteness.  The oscillator could emit or absorb radiation of any
energy.  This was the first parting of ways with classical physics.
Planck however took this assumption as a purely a formal one and did
not quite realise that something radical has been introduced.  As he
said ``This was a purely formal assumption and I really did not give
it much thought except that no matter what the cost, I must bring out
a positive result''.

\subsection{Einstein's light quantum hypothesis}

Einstein was the first person to realise that Planck's introduction of
energy quanta was a revolutionary step.  As we noted a while ago
Einstein, in his light quantum paper, first showed that the so called
`Rayleigh-Jean's Law' is the unambiguous prediction of classical
physics for the radiation law.  This law not only does not work for
high frequency radiation, it also theoretically suffers from
`ultraviolet catastrophy' (i.e. infinite energy).  This convinced
Einstein that to get the correct radiation law, a break with classical
physics is involved. 

In his quest for the cause of the failure of classical physics
Einstein was guided by his unhappiness with asymmetrical treatment of
matter and radiation in classical physics.  Matter is discrete and
particulate while radiation is continnuous and wave like.  He thus
proposes that radiation is also particle-like just as matter is,
i.e. his ``light quantum'' hypothesis, and is not wave like.  He was of
course fully aware of the successes of wave theory in dealing with the
phenomenon of interference and diffraction of light.  All these
phenomenon, however, need only time averages and for such phenomenon
wave theory probably is indispensible.  It is however concievable that
a basic particle picture on time averaging could produce wave like
behavior.  He sumarised that the particle nature of radiation may show
up in the processes involving the generation and transformation of
light where we deal with instantaneous processes.

Can one adduce any evidence in favour of particle nature of light?
Einstein proceeds to show that a consideration of Wien's radiation
law, valid in the nonclassical regime of large frequencies, does
that.  He calculates the probability $p$ that the monochramatic
radiation of frequency $\nu$, occupying a volume $V_0$, could be
confined to smaller volume $V$ using Wien's law.  The result is
\[
p = (V/V_0)^n \ {\rm with} \ n = E/h\nu
\]
where $E$ is the total energy of the radiation.  This is of the same
form as for a gas of $n$ particles.  From this remarkable similarity,
Einstein concludes ``Monochromatic radiation of low density (within
the range of validity of Wien's radiation formula) behaves
thermodynamically as if it consisted on mutually independent energy
quanta of magnitude $R\beta\nu/N$''.  (In modern notation
$R\beta\nu/N$ reads as $h\nu$).  This is the Einstein's light quantum
hypothesis.  In this picutre ``the energy of light is discontinuously
distributed in space.  $\cdots$ when a light ray is spreading from a
point is not distributed continuously over ever increasing spaces, but
consists of a finite number of energy quanta that are localised in
points in space, move without dividing, and can be absorbed or
generated only as a whole''.

Einstein applied successfully his light quantum hypothesis to other
phenomenon involving generation and transformation of light.  The most
important of these was his treatment of photoelectric effect.  He also
discussed Stokes' rule in photoluminescence and ionisation of gases by
ultraviolet light.

\subsection{The photoelectric effect}

In 1887 Heinrich Hertz observed that the ultraviolet light incident on
metals can cause electric sparks.  In 1989 J.J. Thomson established
that the sparks are due to emission of the electrons.  Phillip Lenard
showed in 1902 that this phenomenon, now called the Photoelectric
effect, showed `not the slightest dependence on the light intensity'
even when it was varied a thousandfold.  He also made a
qualitative observation that photoelectron energies increased with the
increasing light frequency.  The observations of Lenard were hard to
explain on the basis of electromagnetic wave theory of light.  The
wave theory would predict an increase in photoelectron energy with
increasing incident light intensity and no effect due to increase of
frequency of incident light.

On the Einstein's light quantum picture, a light quantum, with energy
$h\nu$, on colliding with an electron in the metal, gives its entire
energy to it.  An electron from the interior of a metal has to do some
work, $W$, to escape from the interior to the surface.  We therefore
get the Einstein photoelectric equation, for the energy of the
electron $E$,
\[
E = h\nu - W.
\]
Of course electron may lose some energy to other atoms before escaping
to the surface, so this expression gives only the maximum of
photo-electron energy which would be observed.  One can see that
Einstein's light quantum picture explains quite naturally the
intensity independence of photoelectron energies and gives a precise
quantitative prediction for its dependence on incident light
frequency.  It also predicts that no photoelectrons would be observed
if $\nu < \nu_0$ where $h\nu_0 = W$.  The effect of increasing light
intensity should be an increase in the number of emitted electrons and
not on their energy.  Abram Pais has called this equation as the
second coming of the Planck's constant.

Robert A. Millikan spent some ten years testing Einstein equation and
he did the most exacting experiments.  He summarized his conclusions
as well as his personal dislike of light quantum concept, as follows:
`Einstein's photoelectric equation $\cdots$ appears in every case to
predict exactly the observed results $\cdots$ yet the semi-corpuscular
theory by which Einstein arrived at his equations seems at present
wholly untenable' (1915) and `the bold, not to say reckless hypothesis
of electromagnetic light corpuscle' (1916).

\subsection{Envoi}

Einstein was awarded Nobel Prize in Physics for 1921 for this paper on
light quanta and especially it's application to the photoelectric
effect.  Even though his status as public icon is associated with his
relativity theory, he was not awarded Nobel Prize, for that.  He
however delivered his Nobel Lecture on Relativity.

Einstein's light quantum was renamed as ``photon'' by G.N. Lewis as late
as 1926.  Though Einstein talked about photon energy, $E = h\nu$, in
1905, it is curious that he introduced the concept of photon momentum,
$p = {h\nu \over c}$, only in 1917.  As we have seen even Millikan did
not believe in photons around 1915-16 despite his detailed
experimental work on photoelectric effect.  In 1923, the kinematics of
the Compton effect was worked out on the basis of it's being an
elastic electron-photon scattering by A.H. Compton sucessfully.  After
that it was widely accepted that light does sometimes behaves as
photon.

Einstein made the first application of quantum ideas to matter in his
work on specific heat of solids in 1907.  A consideration of energy
fluctuations, using Planck's radiation law, led him to the dual
particle-wave nature of radiation in 1909.  In 1916-17, in the course
of a new derivation of the Planck's radiation law, using chemical
kinetics methods, Einstein discovered the phenomenon of stimulated
emission of light and introduced his famous $A$ and $B$ coefficients.
These are of fundamental importance in the theory of lasers.

S.N. Bose in 1924 at Dacca sent Einstein a new derivation of Planck's
law in which only the photon concept was used.  Albeit the photons did
not obey the classical statistics of Maxwell and Boltzmann, but rather
a new statistics.  Einstein saw the importance of this contribution,
translated the paper into German, and got it published.  He also
applied it to the matter.  The new statistics is now known as either
Bose-Statistics or as Bose-Einstein statistics.  The particles which
obey this statistics are known as bosons.  As a consequence of these
statistical considerations Einstein discovered that a free gas of Boson
undergoes a phase transition, Bose-Einstein condensation, below a
critical temperature.  The phenomenon was seen only in 1995 and a
Nobel Prize awarded for it in 2001.

The modern mathematical formulation of quantum mechanics was obtained
by W. Heisenberg and E. Schrodinger in 1925-26 in two different, but
equivalent, forms.  Einstein's role in achieving this transition from
old quantum theory to the modern quantum mechanics was quite
significant.  He has been called godfather of Schrodinger's wave
mechanics and his relativity theory with it's emphasis on operational
procedure provided the inspiration to Heisenberg in his matrix
mechanics.

After 1926 Einstein's focus shifted to foundational questions of
quantum mechanics.  He gave his ensemble interpretation of quantum
mechanics.  His discovery of nonlocal correlations in quantum
mechanics with Podolsky and Rosen in 1935 was of far reaching
significance and continues to spawn new fields, such as quantum
computing, quantum information theory and quantum cryplography, down
to the present time.

\section{Epilogue}

Einstein purchased in 1935 a white simple frame house at 112, Mercer
Street within walking distance of his office at Institute for Advanced
Princeton and he lived here till the end.  There were a few pictures
and etchings.  These included a drawing of Gandhi, whom he admired
greatly.  There were photographs of his mother and his sister Maja who
lived with him after she moved to Princeton from Italy in 1939.  He
had also brought with him, from Europe, three etchings of the
physicist he admired more than any other.  These were of Newton,
Maxwell and Faraday.  It is now given his contributions to physics 
abandantly clear that he himself belong to this select company.

When Einstein died the famous cartoonist Herblock published in
Washington Post, a cartoon, in which the planet earth is identified by
the words ``Albert Einstein lived here''.  He became a world icon in
1919 and since then he has continued to hold a place of high esteem in
public mind for his science, his humanity, his fight for a peaceful
world and his freedom from cant.  At the end of millenium He was voted
by Time magazine, and many others as the ``Man of the Millenium''.

\section{Bibiliographical Notes:}

\begin{enumerate}
\item[{1.}] The literature on science and life of Einstein is
enormous.  The best biography for physicists is Pais, A.,
\underbar{Subtle is the Lord $\cdots$: The Science and} \underbar{Life
of Albert Einstein}, Clarendon Press, Oxford and Oxford University
Press, New York, 1982. \\ A volume for a more general reader is
Bernstein, J., \underbar{Einstein}, The Viking Press, New York, 1973. 

\item[{2.}] For the writings of Einstein, we have the multivolume
ongoing series, \\ \underbar{The Collected Papers of Albert Einstein},
Princeton University Press, Princeton, N.J., 1987 -- $\cdots$, and the
companion volumes \\ \underbar{The Collected Papers of Albert
Einstein: English Translation}, Princeton University Press, Princeton,
N.J., 1987 -- $\cdots$.  The brief quotations from Einstein's papers
and letters and thesis reports used in this lecture are from
\underbar{Vol.2: The Swiss Years: Writings, 1900-1909; Vol.5: The
Swiss Years:} \underbar{correspondence, 1902-1914} of this series of
translation volumes. Einstein's papers from the miracle year 1905 are also
available in English translation in \underbar{Einstein's Miraculous
Year: Five Papers} \underbar{that changed the Face of Physics},
(ed. J. Stachel), 
Princeton, 1998.  (Indian reprint by Srishti Publishers, New Delhi,
2001). \\ There are also a number of other english translations of
individual papers.
\item[{3.}] The quote from Isaac Newton is from R.S. Westfall, \\
\underbar{Never at Rest, A Biography of Isaac Newton}, Cambridge,
1981. 
\item[{4.}] On Einstein's doctoral thesis, see also N. Straumann:
arXiv: physics/0504201 (April 2005). 
\item[{5.}] For an appreciation of Einstein's contribution to the
theory of random processes, see L. Cohen, The History of Noise, IEEE
Signal Processing Magazine, p.20-45, Nov. 2005.
\item[{6.}] Ostwald's quote on the reality of atoms is from
Bernstein's book (p.186) referred earlier.
\item[{7.}] For a more detailed writeup on Einstein's contributions to
quantum theory, see V. Singh: Einstein and the Quantum, Current
Science B9, 2101-2112 (2005). \\ 
$[$There is some overlap of the present writeup with this paper$]$
\item[{8.}] For the impact of Einstein's work on the Physics of the
twentieth century, see Physics World, Special Einstein issue,
Jan. 2005; Current Science, Special Einstein Issue, (25 Dec. 2005). \\
D. Giulini and N. Straumann, \underbar{Stud. Hist. Phil. Mod. Phys.},
\underbar{37}, 115-173 (2006).
\end{enumerate}

\end{document}